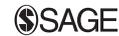

# Are the dead taking over Facebook? A Big Data approach to the future of death online

Carl J Öhman[1]  and David Watson[2]

## Abstract
We project the future accumulation of profiles belonging to deceased Facebook users. Our analysis suggests that a minimum of 1.4 billion users will pass away before 2100 if Facebook ceases to attract new users as of 2018. If the network continues expanding at current rates, however, this number will exceed 4.9 billion. In both cases, a majority of the profiles will belong to non-Western users. In discussing our findings, we draw on the emerging scholarship on digital preservation and stress the challenges arising from curating the profiles of the deceased. We argue that an exclusively commercial approach to data preservation poses important ethical and political risks that demand urgent consideration. We call for a scalable, sustainable, and dignified curation model that incorporates the interests of multiple stakeholders.

> We, the Party, control all records, and we control all memories. Then we control the past, do we not? (Orwell, 1949: 313)

## Introduction

Internet users leave vast volumes of online data behind when passing away, commonly referred to as *digital remains* (Lingel, 2013). The phenomenon is gaining increasing traction within the academic community (Gotved, 2014). Scholars of law and related areas are investigating new dilemmas arising from inheritance of digital estates (Banta et al., 2015; Craig et al., 2013) and issues of posthumous online privacy (Harbinja, 2014). Sociologists and anthropologists are increasingly turning their gaze towards the new types of 'para-social' relationships (Sherlock, 2013), and the 'continuing bonds' (Bell et al., 2015) that we shape with the online dead. And in philosophy, there has been a rising interest for the ontological (Steinhart, 2007; Stokes, 2012; Swan and Howard, 2012) and ethical (Öhman and Floridi, 2018; Stokes, 2015) status of digital remains. In short, online death has rapidly become a booming and diverse research area.

Despite this breadth of perspectives, few studies have thus far explored the macroscopic and quantitative aspects of online death. While research on philosophical micro- and meso-level aspects are illuminating, the global spread of the phenomenon, as well as its future development, remain uncertain. The absence of thorough empirical investigation on the macro-level makes it difficult to formulate a critical analysis of the global impact of online death from either long and/or short-term perspectives. This is problematic, not only because researchers (including the authors of this

[1]Oxford Internet Institute, University of Oxford, Oxford, UK
[2]Oxford Internet Institute, University of Oxford, Oxford, UK; Alan Turing Institute, London, UK

**Corresponding author:**
Carl J Öhman, Oxford Internet Institute, University of Oxford, 1 St Giles, Oxford OX1 3JS, UK.
Email: carl.ohman@oii.ox.ac.uk

study) often motivate the significance of the subject by alluding to its presumed size and growth (Acker and Brubaker, 2014: 10; Harbinja, 2014: 21; Öhman and Floridi, 2017: 640), but also because there is reason to believe that online death will increase in significance as more people around the world become connected and mortality numbers rise. It is important to get the picture straight. Is social media, as occasionally claimed (Ambrosino, 2015; Brown, 2016), turning into a 'digital graveyard'? If so, how is the phenomenon geographically distributed? And perhaps more importantly, what ethical and political challenges would emerge from such development? Despite the somewhat alarming nature of these questions, there have hitherto been few attempts to provide rigorous answers.

To address this lacuna and lay the groundwork for further macroscopic analysis, the current study sets out to estimate the growth of digital remains over the course of the 21st century, using the world's largest platform – Facebook – as a case study. Facebook's policy on deceased users has changed somewhat over the years, but the current approach is to allow next of kin to either memorialize or permanently delete the account of a confirmed deceased user (Facebook, n.d.).[1] The focus of this article, however, is not merely on the memorialized profiles, but on all profiles belonging to deceased users, be they memorialized or not. We pose two research questions:

> RQ1: How will the number of Facebook profiles belonging to dead users develop over the course of the 21st century?
>
> RQ2: What will be the geographical distribution of dead Facebook profiles?

Our analysis is conducted in two stages, henceforth referred to as Scenarios A and B. In Scenario A, we assume a global freeze on new users joining the network as of 2018 and predict the resulting accumulation of dead profiles for each nation in the world. This effectively sets a 'floor' on the possible growth of dead profiles on the network. To carry out the analysis, we use a public dataset of projected mortality from 2000 to 2100, distributed by age group and nationality (United Nations, Department of Economic and Social Affairs, 2017). These data are matched with current Facebook user totals, scraped from Facebook's audience insights application programming interface (API) for each country and age group. This allows us to estimate the number of Facebook users expected to die in any given country-year. In Scenario B, we expand the analysis to a hypothetical scenario for Facebook's future growth, assuming that the network will continue to grow at a pace of 13% per year (Facebook, 2018) until it reaches a penetration rate of 100% for each country-year-age group. While unlikely, this estimate provides a 'ceiling' for the accumulation of dead profiles. In conjunction with the 'floor' defined in Scenario A, this ceiling defines the window within which we can expect the true number of dead profiles to fall.

In concluding the study, we situate the findings within the larger context of digital preservation (Whitt, 2017). We raise concerns over the current dominance of commercial data management, and warn that it may limit future generations' access to historical data. We argue that profiles of the deceased are valuable in ways that cannot be quantified in purely economic terms, which is why we advocate an explicitly multi-stakeholder approach. If data are preserved solely on the basis of corporate profitability, we warn that non-economic considerations – e.g., the ethical, religious, scientific, and historical value of digital remains – may be neglected. Our digital heritage is difficult to measure in dollars and cents.

## Data

Three types of data were used to carry out the analysis: projected mortality over the 21st century, distributed by age and nationality; projected population data over the 21st century, also distributed by age and nationality; and current Facebook user totals for each age group and country.

Mortality rates were calculated based on UN data, which provide the expected number of mortalities and total populations for every country in the world (United Nations, Department of Economic and Social Affairs, 2017). Numbers are available for each age group – 0 to 100, divided into five-year intervals – and all years from 2000 to 2100, likewise divided into five-year intervals. The estimates are based on official data from each country's government, and in some cases external sources (esa.un.org/unpd/wpp/DataSources/). It is unclear from the data how precision varies by country and year. All projections are reported as point estimates, with no standard errors or confidence intervals. For a more detailed account of the UN data, see esa.un.org/unpd/wpp/.

Facebook data were scraped from the company's Audience Insights page (facebook.com/ads/audience-insights/) using a custom Python script that extracts Facebook's active monthly users by country and age. These estimates are based on the self-reported age of users. Facebook provides lower and upper bounds for user totals across all ages and nationalities. For example, there are between 15 and 20 million 25-year-old Indians on the network.[2] Variability increases with user counts, both of which are reported in round numbers divisible by 5 or 10, suggesting that they are not meant as serious estimates of standard errors or confidence intervals. We take the midpoint of each country-age window for our analysis.

Facebook's audience insights API provides by far the most comprehensive publicly available estimate of the network's size and distribution. Nevertheless, we wish to draw attention to several limitations of this dataset. First, there are reasonable doubts about the accuracy of Facebook's reported monthly active users. The site has recently been sued for allegedly inflating these numbers with the intent of overcharging advertisers (Todd, 2018), and Facebook explicitly notes that their estimates are not meant to be matched with population data. In addition to these concerns about false positives, we also expect false negatives due to users visiting the site less than once a month. Unfortunately, it is impossible to say exactly how this affects results without more fine-grained detail on the distribution of errors. Second, the data exclude users under 18, preventing us from evaluating network activity among 13–17 year olds (Facebook requires all users to be at least 13). Due to the relatively low (although varied) mortality rate of this age group, the missing data should not have much impact on the projection until relatively late in the century. Third, users aged 65+ are all put into the same age category. This gives us less detailed data on penetration rates among the elderly. But as we show in the following section, this problem can be mitigated by extrapolating from a smooth curve fit to data from younger users.

Finally, we wish to emphasize that our model is devoted to the future development of death on Facebook, and therefore leaves out users who have already died and left profiles behind. Estimating the current number of dead profiles would require historical data on the age distribution of Facebook users in various countries, which are currently inaccessible through the site's API. Furthermore, the aim of the study is to depict a larger, long-term trend, in which the current numbers play only an illustrative role.

## Methodology

Our methodological approach can be summarized by the following procedure for each country:

1. estimate a function $f$ mapping age and year to expected mortality rates (see Figure 1(a));
2. estimate a function $g$ mapping age to expected active monthly Facebook users (see Figure 1(b));
3. extend $g$ across time under two alternative scenarios (details below);
4. multiply the outputs of $f$ and $g$ to estimate the number of Facebook profiles belonging to dead users of a given age in a given year (see Figure 1(c)); and
5. integrate this product across all age groups to estimate the number of dead profiles in a given year.

This pipeline is repeated for each country to get a global estimate. Projections are integrated over several years to get national or global estimates over time.

It should be noted that this approach makes a substantive and potentially problematic assumption, namely that each country's Facebook users constitute a representative sample of the population, at least with respect to mortality rates. It is well established that internet usage, especially in developing economies, is strongly correlated with education and income (PEW Research Centre, 2018: 15). These two variables are in turn correlated with life expectancy, which means there is reason to believe that current Facebook users will live slightly longer than non-users on average. Our model does not account for this potential bias, which may result in an overestimation of dead users in developing countries.

However, a recent PEW research report (2018: 15) indicates that the divide is rapidly shrinking. Between 2015 and 2017, social media penetration in countries such as Lebanon, Jordan and the Philippines rose by more than 20 percentage points, suggesting that connectivity is fast becoming increasingly accessible. This trend is expected to continue throughout the 21st century, mitigating any potential confounding effects on projections years or decades out. Furthermore, the closer we get to full market saturation, the smaller the bias becomes since people with high and low life expectancies are both joining the network in large numbers. In the face of this, it is important to stress that the value of the present study lies in the larger trends it identifies, not in the details of the immediate future development. This should be kept in mind when assessing very short-term scenarios.

The model described in step (2) was trained on 2018 data. We vary projections for future Facebook growth according to two scenarios: (A) *Shrinking*. No new users join the network. All current users remain until their death. (B) *Growing*. The network grows at 13% per year across all markets until usership reaches 100%. To help extrapolate beyond the age of 64, the final age for which Facebook provides monthly active user totals, we anchored all regressions with an extra data point of zero users aged 100. This is almost certainly true in all markets, at least to a first approximation. Alternative anchor points may be justified, but do not have a major impact on results.

All statistical analysis was conducted in R, version 3.5.1 (R Core Team, 2018). Predictive functions were estimated using generalized additive models (GAMs), which provide a remarkably flexible framework for learning nonlinear smooths under a wide range of settings (Hastie and Tibshirani, 1990). Regressions were implemented using the mgcv package (Wood, 2017). A supplemental methods section, including data and code for reproducing all figures and results, can be found online at: https://github.com/dswatson/digital_graveyard.

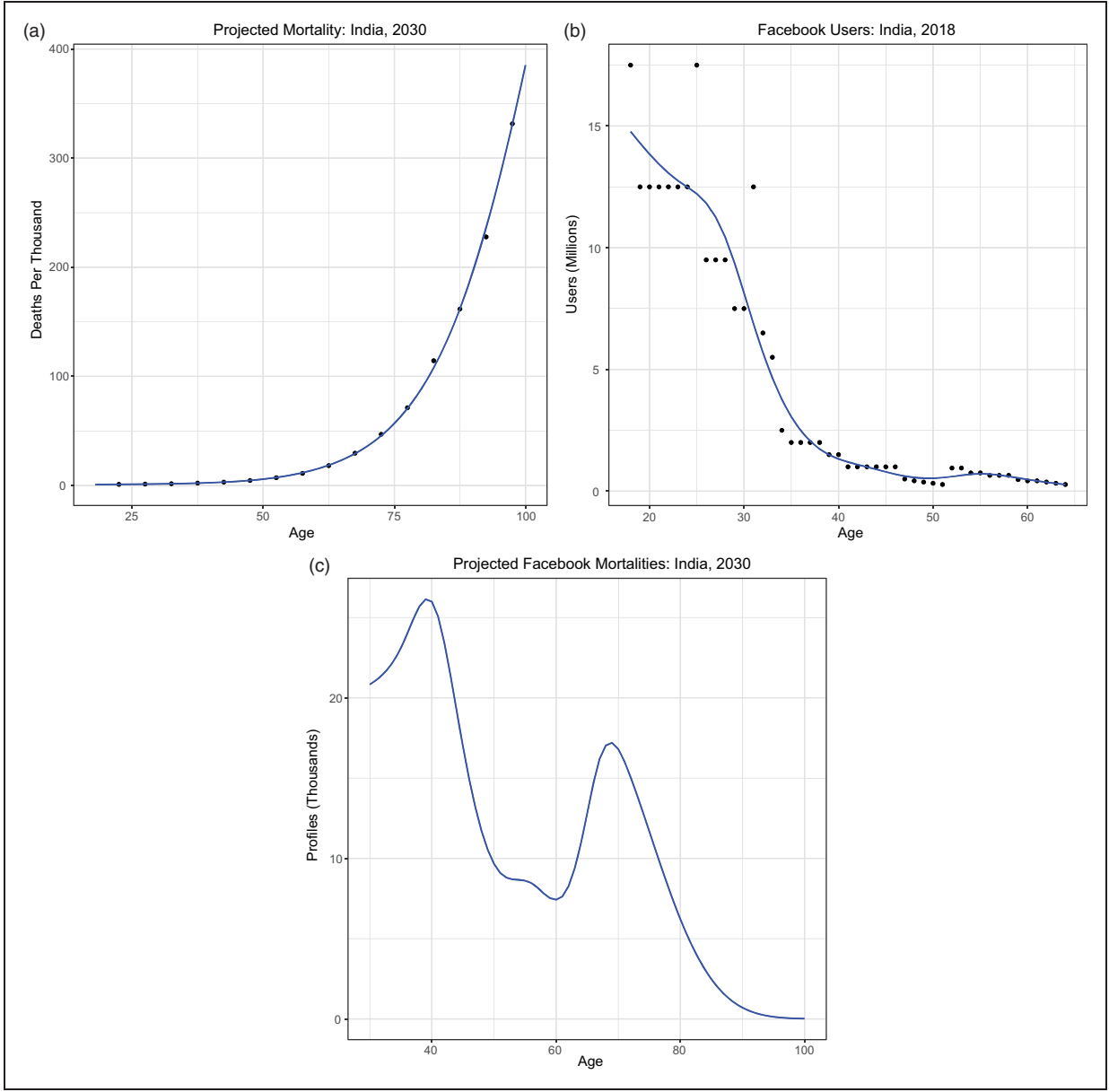

**Figure 1.** The analysis pipeline under Scenario A for India 2030. (a) Mortality rate is modelled as a function of age. (b) Facebook user totals as a function of age. (c) Predicted values for both functions are multiplied to estimate the number of dead Indian users. The area under this curve is our projected number of Indians on Facebook who will die in 2030.

We fit three separate models for each country

$$\text{Mortality\_Rate} = f_C(\text{Time}, \text{Age})$$
$$\text{FB\_Users\_2018} = g_C(\text{Time} = 2018, \text{Age})$$
$$\text{Population} = h_C(\text{Time}, \text{Age})$$

The subscript $C$ indicates that each model is country-specific. We omit the subscript for notational convenience moving forward.

The mortality and population models provide nonlinear interpolations so that we can make predictions for any age-year in the data without the limitations imposed by the UN's binning strategy.

Under Scenario A, we extrapolate model $g$ beyond 2018 by assuming that no new users join Facebook and current users leave the network if and only if they die. This means we see zero 18-year-olds on the network in 2019, zero 18- or 19-year-olds in 2020, and so on.

Attrition from current users can be calculated recursively. For each year $t$ and age $a$:

Scenario A

$$\text{FB\_Users} = g(\text{Time} = t, \text{Age} = a) =$$
$$g(\text{Time} = t-1, \text{Age} = a-1)$$
$$\times (1 - f(\text{Time} = t-1, \text{Age} = a-1))$$

In Scenario B, we extrapolate beyond $g$ by assuming that Facebook will see constant growth of 13% per year in all markets until reaching a cap of 100% penetration. For each year $t$ and age $a$:

Scenario B

$$\text{upper\_bound} = h(\text{Time} = t, \text{Age} = a)$$
$$\text{FB\_proj} = g(\text{Time} = t-1, \text{Age} = a-1) \times 1.13^{t-2018}$$
$$\text{FB\_Users} = g(\text{Time} = t, \text{Age} = a)$$
$$= \min(\text{upper\_bound}, \text{FB\_proj})$$

In both cases, our true target is

$$y = \int_{13}^{100} \int_{2018}^{2100} f(\text{Age}, \text{Time}) g(\text{Age}, \text{Time}) d(\text{Age}) d(\text{Time})$$

For the mortality rate model $f$, we used beta regression with a logit link function, a common choice for rate data. For the Facebook model $g$, we used negative binomial regression with a log link function, which is well suited for over-dispersed counts such as those observed in this dataset. We experimented with several alternatives for the population model $h$, ultimately getting the best results using Gaussian regression with a log link function. Parametric specifications for each model were evaluated using the Akaike information criterion (Akaike, 1974), a penalized likelihood measure. Age and time were incorporated as both main effects and interacting variables in models $f$ and $h$, which were fit with tensor product interactions in a functional ANOVA structure (Wood, 2006). We use cubic regression splines for all smooths, with a maximum basis dimension of 10. Parameters were estimated using generalized cross-validation.

## Uncertainty

While there remains no good way to evaluate the precision of the underlying data – as noted above, neither the UN nor Facebook provides confidence intervals – we may quantify the uncertainty of the model using nonparametric techniques. GAMs provide straightforward standard errors for their predictions, but under both scenarios our true target $y$ is a double integral of a product of two vectors. Unfortunately, there is no analytic method for calculating $y$'s variance as a function of those variables without making strong assumptions that almost certainly fail in this case.

For that reason, we measure uncertainty using a Bayesian bootstrap (Rubin, 1981). To implement this algorithm, we sample $n$ weights from a flat Dirichlet prior and fit the models using these random weights. We repeat this procedure 500 times for each country and scenario, providing an approximate posterior distribution for all predictions, from which we compute standard errors. These numbers are reported in parentheses next to point estimates in the text, and in their own column in all table summaries.

## Findings

As previously noted, the findings we present in this paper concern only the *future* accumulation of dead profiles (i.e., those who will die between 2018 and 2100). Naturally, many users have already left profiles behind when they passed away. This number, however, is unknown, but should (whatever it is) be added to the plots we present in Scenarios A and B below.

### Scenario A

Our first scenario assumes that users will cease joining the network as of 2018. While unlikely, this defines the minimum of the possible development, what we refer to as the floor (see Figure 2). Attached to the plot is a table with the exact numbers and share of each continent (Table 1).

Under the assumptions of Scenario A, we estimate that some 1.4 billion (±11.15 million) Facebook users will die between 2018 and 2100 – fully 98% of the 1.43 billion users in our dataset. Under this scenario, the number of deaths per year on Facebook grows steadily for the next five decades, peaking at over 29 million (±0.31 million) in 2077 before decelerating through the rest of the century. The global sum of dead profiles exceeds 500 million (±3.86 million) in 2060 and 1 billion (±8.67 million) in 2079. Note that under these conservative assumptions, the dead will in fact overtake the living on Facebook in about 50 years. This corroborates popular claims in media (Ambrosino, 2015; Brown, 2013) about living profiles becoming a minority on the network within the (relatively) near future.

The plot further shows that Asia contains a growing plurality of deceased users for every year in the dataset, culminating with nearly 44% of the total by the end of the century. Nearly half of those profiles come from just two countries, India and Indonesia, which account for a cumulative 278.8 million (±9.8 million) Facebook mortalities by 2100 (see Table 2).

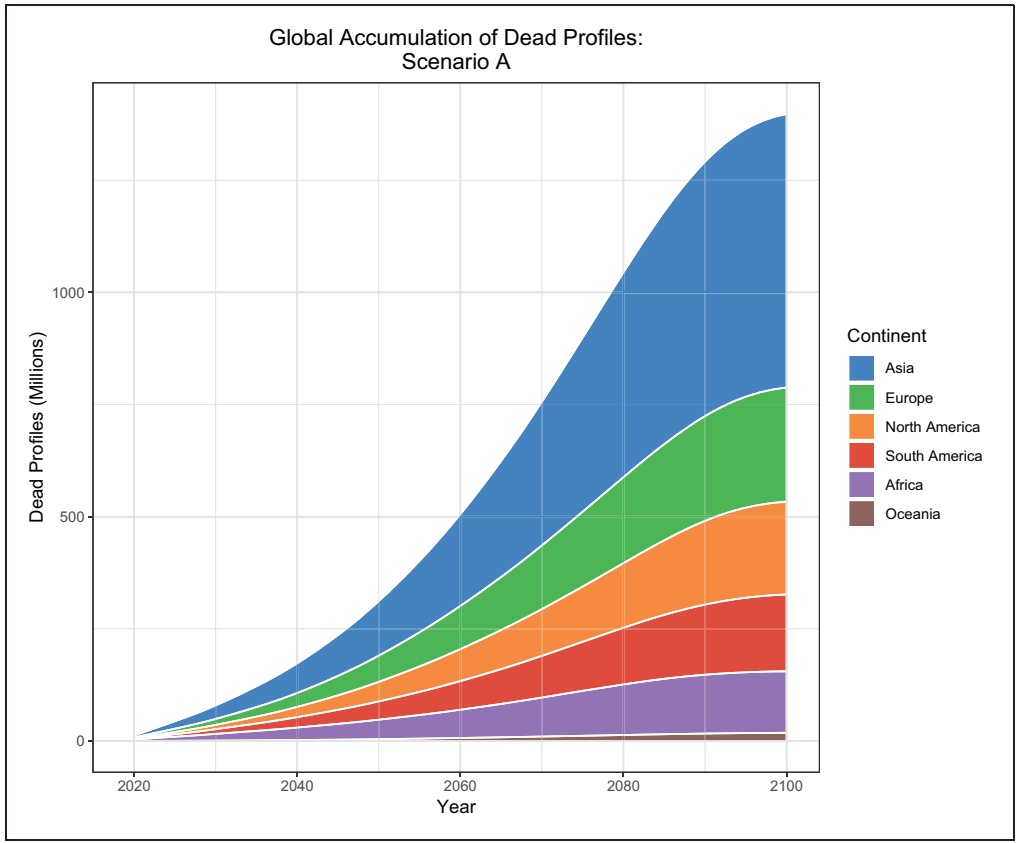

**Figure 2.** Accumulation of dead profiles under Scenario A.

**Table 1.** Geographical distribution of dead profiles (in millions) under Scenario A.

| Time | Continent | Profiles | SE | Percentage |
|---|---|---|---|---|
| 2100 | Africa | 137.3577 | 1.0876 | 9.8273 |
| 2100 | Asia | 609.7714 | 10.1823 | 43.6261 |
| 2100 | Europe | 254.3236 | 1.1467 | 18.1956 |
| 2100 | North America | 206.9839 | 2.3485 | 14.8087 |
| 2100 | Oceania | 18.1326 | 0.1467 | 1.2973 |
| 2100 | South America | 171.1529 | 3.3918 | 12.2451 |

**Table 2.** Geographical distribution of dead profiles (in millions) by country under Scenario A. Results for top ten countries shown.

| Time | Country | Profiles | SE | Percent |
|---|---|---|---|---|
| 2100 | India | 207.6545 | 8.3233 | 14.8527 |
| 2100 | United States | 115.7516 | 2.0561 | 8.2793 |
| 2100 | Indonesia | 71.1468 | 5.2658 | 5.0889 |
| 2100 | Brazil | 65.3675 | 3.3203 | 4.6755 |
| 2100 | Mexico | 42.9421 | 1.0875 | 3.0715 |
| 2100 | Philippines | 34.6779 | 0.7469 | 2.4804 |
| 2100 | United Kingdom | 31.7917 | 0.3797 | 2.2739 |
| 2100 | France | 29.7236 | 0.2221 | 2.1260 |
| 2100 | Thailand | 28.9506 | 0.8205 | 2.0707 |
| 2100 | Vietnam | 28.5697 | 0.6811 | 2.0435 |
| 2100 | Rest of World | 741.5138 | 2.3780 | 53.0376 |

## Scenario B

Scenario A is highly unlikely. For Facebook to see zero global growth as of 2019 would require some cataclysmic event(s) far more ruinous than the Cambridge Analytica scandal (Cadwalladr and Graham-Harrison, 2018), which revealed serious issues regarding the security and privacy of Facebook user data. To estimate how much higher the growth can possibly be, the second scenario sets a 'ceiling' on the development. We presume that Facebook will continue to see global growth of 13% per year until it reaches 100% penetration in all markets. As illustrated by Figure 3, this assumption drastically changes the total number of dead users by the end of the century.

A continuous growth rate of 13% per year increases the expected number of dead profiles on Facebook by a factor of 3.5, for a total sum of 4.9 billion (±97.23 million). Unlike Scenario A, the dead profiles do not show any signs of exceeding the living within this century. However, the proportion is still substantial, and the dead are likely to reach parity with the living in the first decades of the 22nd century.

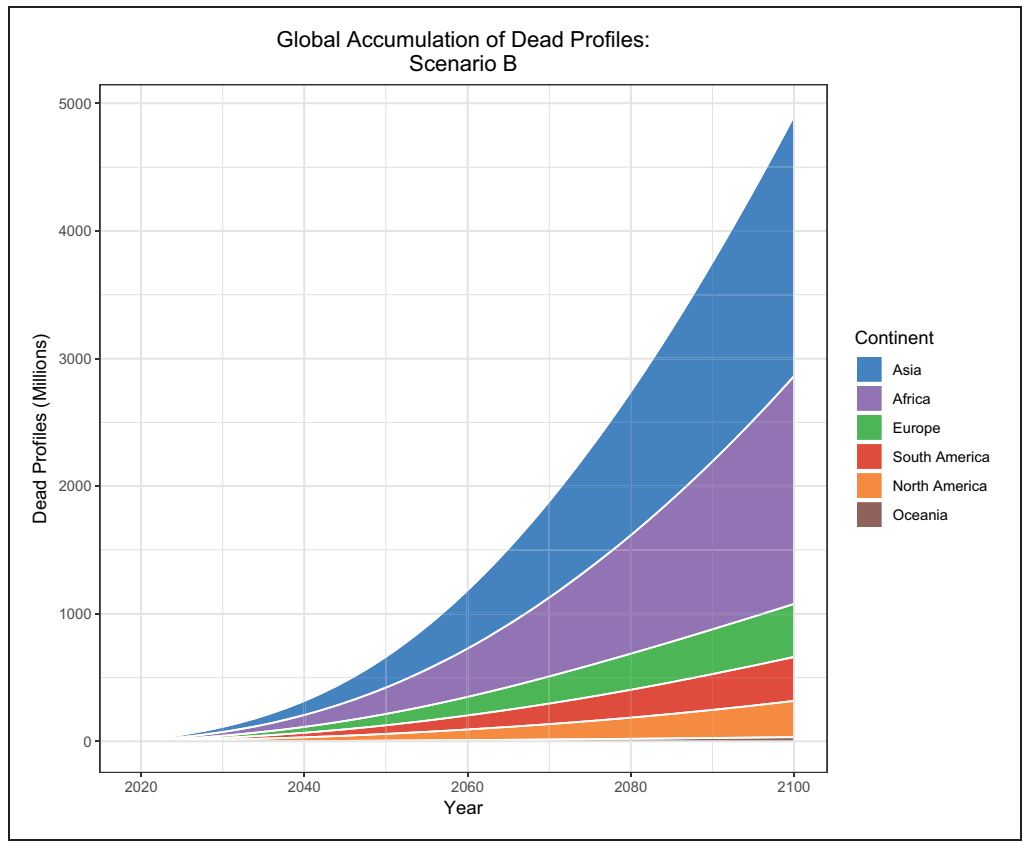

**Figure 3.** Accumulation of dead profiles under Scenario B.

**Table 3.** Geographical distribution of dead profiles (in millions) under Scenario B.

| Time | Continent | Profiles | SE | Percentage |
| --- | --- | --- | --- | --- |
| 2100 | Africa | 1786.4705 | 73.2377 | 36.4435 |
| 2100 | Asia | 2039.0000 | 40.5229 | 41.5951 |
| 2100 | Europe | 414.7771 | 9.3840 | 8.4613 |
| 2100 | North America | 283.6859 | 33.2734 | 5.7871 |
| 2100 | Oceania | 33.6563 | 1.1270 | 0.6866 |
| 2100 | South America | 344.4356 | 31.8269 | 7.0264 |

**Table 4.** Geographical distribution of dead profiles (in millions) by country under Scenario B. Results for top ten countries shown.

| Time | Country | Profiles | SE | Percent |
| --- | --- | --- | --- | --- |
| 2100 | India | 783.7010 | 30.6525 | 15.9852 |
| 2100 | Nigeria | 315.5320 | 39.8116 | 6.4359 |
| 2100 | Indonesia | 221.0466 | 21.1954 | 4.5087 |
| 2100 | Pakistan | 177.0034 | 3.3463 | 3.6104 |
| 2100 | Brazil | 144.7937 | 31.7306 | 2.9534 |
| 2100 | Niger | 126.7529 | 29.0875 | 2.5854 |
| 2100 | United States | 112.4368 | 28.9811 | 2.2934 |
| 2100 | Philippines | 102.4860 | 2.8522 | 2.0904 |
| 2100 | Mali | 100.0100 | 17.3816 | 2.0399 |
| 2100 | Burkina Faso | 93.5536 | 26.9056 | 1.9082 |
| 2100 | Rest of World | 2725.3420 | 56.1275 | 55.5891 |

A continuous 13% growth rate would change not just the total number of dead users, but their geographical distribution (see Tables 3 and 4). The most notable shift is the considerably increased share of global Facebook mortalities contributed by African nations. Nigeria in particular becomes a major hub of Facebook user deaths under Scenario B – in fact, the second largest in the world, accounting for over 6% of the global total. The shift is evident in Figures 3 and 4. Niger, Mali and Burkina Faso also appear in the top 10 countries by dead profile count, while the United States is the only Western nation to crack the list. In other words, a minority of dead profiles will belong to Western users.

To illustrate the geographical distribution more clearly, we have included a heatmap that visualizes deceased Facebook users per country (see Figure 4). Unsurprisingly, the map closely tracks the list of largest Facebook markets. However, it should be noted that only two Western countries (the US and the UK) make it to the top 10 list under either scenario. Thus, the maps clearly show that death online is a global phenomenon, reaching far wider than just Europe and America.

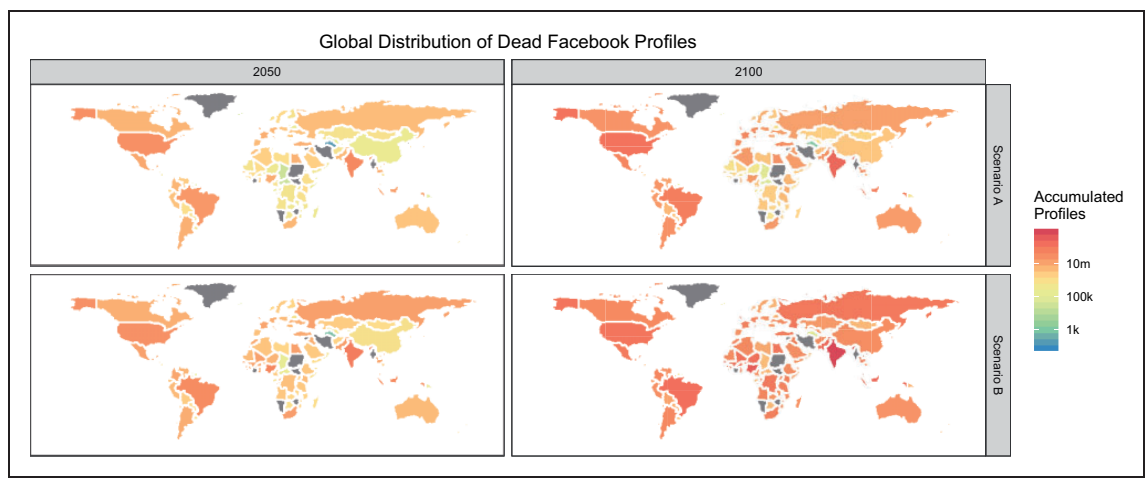

**Figure 4.** Heat map visualizing the global distribution of deceased Facebook user profiles under Scenarios A and B. Numbers are plotted on a logarithmic scale. Countries and regions with no Facebook data or fewer than 10,000 monthly active users were not included in our models and are rendered in grey.

To summarize, both scenarios are implausible. The true number almost certainly falls somewhere between Scenarios A and B, but we can only speculate as to where. Assumptions regarding growth rates have a major impact on both absolute numbers and geographical distributions of dead profiles. While richer data sources may help produce more accurate projections, an exact estimate is almost beside the point. Even in the conservative Scenario A, numbers are large. Facebook will indubitably have hundreds of millions of dead users by 2060 if not sooner.

With regards to the geographical distribution, it can be noted that in both scenarios, a handful of countries make up a large proportion of the total – mainly India (due to its large population) and the US (due to its high penetration rates), but also other countries like Nigeria and Brazil will be important stakeholders in this development. Next, we turn to a discussion of the challenges posed by the growth of death online.

## Discussion

Our projection of growth in dead Facebook users' accounts marks the first step toward empirically exploring the macroscopic and quantitative aspects of death on social media. The results should be interpreted not as a prediction of the future, but as a commentary on the present, and an opportunity to respond with thoughtful and effective policy interventions.

Undoubtedly, there is a great deal of uncertainty in projections of this kind. In addition to the predictive variance discussed above, there is also uncertainty regarding the data underlying the model. For instance, we do not know if there will be a significant cultural shift among users towards deleting profiles (either one's own or deceased relatives', a possibility given to the appointed legacy contact). It is also possible that Facebook will unexpectedly go bankrupt in the foreseeable future, thus invalidating the assumptions underlying our models. As stressed by boyd (2006) among others, the longevity of social media sites depends on their ability to evolve, and despite the success of the past decade, we do not yet know how or if Facebook will manage to do this in the future.

But this has no bearing on our larger point – namely, that critical discussion of online death and its macroscopic implications is urgently needed (not least in regard to its geographical spread). Facebook is merely an example of what awaits any platform with similar connectivity and global reach. Furthermore, the sudden dissolution of Facebook would arguably make the subject even more important, as the company may be forced to sell or delete their user data. A sufficiently severe blow to Facebook's finances could force a redesign of the platform with major implications for those currently using it as a memorial site (see, for instance, Arnold et al., 2018: 202 on how the relaunch of MySpace 2013 dropped features used by mourners). In what follows, we tentatively presume that Facebook or something like it will continue to exist for the foreseeable future.

Each individual who leaves a profile behind represents a unique event in its own right, which often leaves us with difficult questions of inheritance of digital assets (Banta et al., 2015; Craig et al., 2013) and posthumous online privacy (Harbinja, 2014). But when aggregated, the totality of these cases amounts to something beyond the sum of its parts. The personal digital heritage left by the online dead are, or will at least become, part of our shared *cultural* digital heritage (Cameron and Kenderdine, 2007), which may prove invaluable not only to future historians (Brügger and Schroeder, 2017; Pitsillides et al., 2012; Roland and Bawden, 2012), but to future generations as part of their record and self-understanding. As stated by Matt Raymond, the former director of communications at the American Library of Congress upon receiving a large data donation from Twitter, 'Individually tweets might seem insignificant, but viewed in the aggregate, they can be a resource for future generations to understand life in the 21st century' (Raymond, 2010). Such records can thus be thought of as a form of future *public good* (Waters, 2002: 83), without which we risk falling into a 'digital dark age' (Kuny, 1998; Smit et al., 2011).

Despite its seeming immortality, digital information is more fragile than is sometimes assumed, and future access is far from guaranteed (Whitt, 2017) – even for Facebook itself. File formats change, hardware must be updated, and data need to be continuously stewarded and organized in order to remain useful. As Jeff Rothenberg (1995) says, 'Digital information lasts forever – or five years, whichever comes first.' This is not primarily due to storage costs. 'The real cost of storage', as Palm (2006: 5) puts it, 'is management.' To maintain data utility, firms must routinely upgrade systems and tend to their contents, a costly and tedious undertaking for which Facebook's current curation model was not designed. Lavoie and Dempsey (2004: 229) put it well:

> Preserving our digital heritage is more than just a technical process of perpetuating digital signals over long periods of time. It is also a social and cultural process, in the sense of selecting what materials should be preserved, and in what form; it is an economic process, in the sense of matching limited means with ambitious objectives; it is a legal process, in the sense of defining what rights and privileges are needed to support maintenance of a permanent scholarship and cultural record...And perhaps most importantly, it is an ongoing, long-term commitment, often shared, and cooperatively met, by many stakeholders.

While Lavoie and Dempsey write primarily for an audience of librarians and archivists, their argument is equally applicable to the case of digital remains on Facebook: the cultural/ethical process of selecting whose data are worth preserving, and *how* to preserve them, is inseparable from the economic constraints that induce the question. But how is one to determine what is worth preserving? This requires a normative framework, one or several guiding principles that help us determine the value of data. There are many possible candidates for such a principle. An object can be appreciated for its sentimental, scientific, religious, or aesthetic values, to list just a few considerations of note. Furthermore, it is plausible that different regions, nations and other interest groups will appreciate different values in Facebook's mounting historical record. Nevertheless, it is neither users nor their political representatives or religious groups who determine how their data is collectively managed – it is the corporate interests of Facebook.

For a firm, what makes data 'worth preserving' is ultimately their ability to directly or indirectly contribute to the company's profit. Data belonging to deceased users may prove valuable for such purposes. For example, the memorialized profiles may still serve the function of attracting living users who visit the profile to mourn (Karppi, 2013). Indeed, time spent on Facebook can even be understood as a type of labour (Fuchs and Sevignani, 2013). While the (indirect) traffic generated by mourning relatives may not single-handedly result in enough clicks and exposure to cover the costs of curating the dead, it could still serve the indirect function of appropriating central social functions such as mourning and love (Öhman and Floridi, 2017). What is more, datasets of digital remains may also be used for training new models (Leaver, 2013) and extracting historical insight, which may provide a valuable market advantage. Few legal obstacles stand in the way of such experimentation, as deceased users are not, at least according to current legislation, protected the way living users are (see, for instance, the latest GDPR, which lacks any clear guidelines for handling digital remains).

While both the traffic generated by the bereaved and the internal training of new models are possible uses of digital remains, they do not guarantee long-term profitability. If the economic value of dead profiles were ever to become negative, market forces would compel a rationally self-interested firm to delete them. This seems to be the preferred option for many other social media including Twitter (Twitter.com, n.d.), and is also advocated on a normative basis by some scholars, perhaps most notably Mayer-Schönberger (2009). But thus far Facebook appears to have found the net value of dead profiles to be positive.

This is not to say that Facebook, nor any other platform, only appreciates the commercial value of digital remains as a source of financial exploitation. In fact, Facebook has carefully considered the ethical implications of their policy (Brubaker and Callison-Burch, 2016), and has removed advertisements from the deceased profiles, thus virtually de-commercializing the space. But the de-commercialization itself may be interpreted as a response to market incentives, in so much as it is rational for firms to maintain the good will of their customers. Curating a deceased relative's profile could keep some users on the platform, even if it is not the main source of revenue generated by them.

Market incentives may often overlap with the interests of researchers, consumers and future generations – but they are by no means identical. Markets have been discussed rather extensively in the digital preservation literature. For instance, Lavoie (2003: 15) identifies three ideal type-roles in the economics of digital preservation: Right holder, Archive and Beneficiary. Sometimes, these roles are played by a single entity, sometimes by separate ones. Lavoie stresses that in so-called supply-side models (17), where the Right holder and the Archive are the same entity but the Beneficiary is external, there is a risk that the market does not create sufficient incentives for preservation. This is indeed a risk in the case of Facebook. The platform has both the rights to the information stored and is the archiving entity. Moreover, they have little incentive to share (to say nothing of the complexity of posthumous privacy rights). The beneficiaries – in this case future generations and historians – can neither speak for themselves nor create any current incentives, which make a purely free-market model inappropriate.

This situation requires what one may call *a new macro-ethics of deletion* (to borrow a term from Floridi, 2013), a curation model that encompasses and appreciates the various kinds of values involved. In line with Lavoie and Dempsey's argument, we therefore conclude that multiple stakeholders must be considered. These stakeholders may include states, NGOs, universities, libraries, museums, and any other kind of institution that provides unique perspectives on the value of our digital heritage. The multi-stakeholder approach is not in itself a novel proposal. Indeed, the pioneering Task Force on Archiving of Digital Information (1996) was composed of a collection of individuals representing industry, museums, archives and libraries, publishers, scholarly societies and government. And newer initiatives, such as UNESCO's strategic plan for software heritage (Di Cosmo and Zacchiroli, 2017: 4), have continued to stress the value of diversity in digital preservation:

> We believe that, for Software Heritage, it is essential to build a not-for-profit foundation that has as its explicit objective the collection, preservation and sharing of our software commons. In order to minimize the risk of having a single point of failure at the institutional level, this foundation needs to be supported by various partners from civil society, academia, industry, and governments, and must provide value to all areas that may take advantage of the existence of the archive, ranging from the preservation of cultural heritage to research, from industry to education.

While the above quote deals mainly with software, the same can be said about the vast datasets accumulated by social media firms. It is important that historically significant data are preserved in a way that serves *all* of humanity, and this cannot be done by allocating the curation of historical social records to any one agent operating in its rational self-interest.

Finally, we wish to stress the importance of decentralizing control over aggregates of digital remains. Concentration of historical data in private hands may prove problematic for political reasons (Lor and Britz, 2012; Öhman, 2018). While it is true that one's digital remains are often distributed over multiple platforms and media (Cann, 2014; Pitsillides et al., 2012: 19), it seems that control of personal data (and hence digital remains) are increasingly concentrated in a small number of global actors (many of which are owned by Facebook, e.g. Whatsapp, Messenger, and Instagram). And, as Orwell so adroitly observed in *1984*, those who control our access to the past also control how we perceive the present. So, in order to prevent a possibly dystopian future of power asymmetries and distorted historical narratives, the task before us is to design a *sustainable*, *dignified* solution that takes into account *multiple* stakeholders and values. This inevitably requires a decentralization of control and ownership of our collective digital heritage.

Academic knowledge will be key in this process. Researchers are charged not just with providing macro-level analyses like this one, but also with providing qualitative knowledge of how individuals in different cultures and social settings make sense of death and the digital. When it comes to qualitative research, there is already a rich literature upon which to draw (Bell et al., 2015; Brubaker et al., 2016; Kasket, 2012). However, researchers have hitherto mainly focused on North American and European settings – with some exceptions (Choudhary, 2018). If the goal is to contribute to a fair and flexible system for curating digital remains, researchers must increasingly turn to non-western contexts, where the phenomenon is going to have the largest presence. While survey data from previous studies do not indicate any radical differences in attitudes toward online death across cultures (Grimm and Chiasson, 2014), a qualitative, nuanced understanding of this fast-evolving subject is required. We therefore encourage scholars of online death to widen the geographical scope of their research, and focus particularly on South Asia and Africa, where our models suggest the phenomenon will be most prevalent in the coming decades.

## Conclusion

This study has provided the first rigorous projection of the accumulation of Facebook profiles belonging to the deceased. Will the dead then, 'take over' Facebook? We have concluded that hundreds of millions of dead profiles will be added to the network in the next few decades alone, and that the dead may well outnumber the living before the end of the century, depending on how global user penetration rates evolve. Irrespective of how the network grows in the years to come, the vast majority of dead profiles will belong to users from non-western countries.

Considering its global reach, we have argued that the totality of deceased user profiles amounts to something beyond the sum of its parts. These profiles are becoming part of our collective record as a species, and may prove invaluable to future generations. We believe that a multi-stakeholder approach is the best way to curate such a vast archive. We have also stressed that in crafting a future curation model, qualitative understanding of how different cultures make sense of death and the digital will be key. Likewise, the development poses difficult ethical problems that require careful consideration. The onus is now on policymakers and industry to rise to these challenges. We look forward to taking part in the debates to come.

### Acknowledgements
We wish to express our sincere gratitude to the four referees who reviewed this study. Their insights and input have substantially improved the final result. We would also like to express our thanks to Patrick Gildersleve for helping us with the Python script that scraped data from the Facebook API.

### Declaration of conflicting interests
The author(s) declared no potential conflicts of interest with respect to the research, authorship, and/or publication of this article.

### Funding
The author(s) received no financial support for the research, authorship, and/or publication of this article.

### Notes
1. Users are also encouraged to select a 'legacy contact' that will steward the account upon their death.
2. For the record, it should be noted that while Facebook is popular in most countries, it faces considerable competition from for example VKontakte in Russia, and is almost completely absent in other places (China, North Korea, etc.). This, however, is not our main concern.

### ORCID iD
Carl J Öhman  http://orcid.org/0000-0002-8887-5324
David Watson  http://orcid.org/0000-0001-9632-2159

### References

Acker A and Brubaker JR (2014) Death, memorialization, and social media: A platform perspective for personal archives. *Archivaria* 77: 2–23.

Akaike H (1974) A new look at the statistical model identification. *IEEE Transactions on Automatic Control* 19(6): 716–723. http://doi.org/10.1109/TAC.1974.1100705.

Ambrosino B (2015) Facebook as a growing and unstoppable graveyard. In: *BBC.com*, 14 March. Available at: http://www.bbc.com/future/story/20160313-the-unstoppable-rise-of-the-facebook-dead (accessed 1 August 2016).

Arnold M, Gibbs M, Kohn T, et al. (2018) *Death and Digital Media*. London: Routledge.

Banta BNM, Jacob BR, Assistant V, et al. (2015) The role of private contracts in distributing or deleting digital assets at death. *Fordham Law Review* 83.

Bell J, Bailey L and Kennedy D (2015) We do it to keep him alive': Bereaved individuals' experiences of online suicide memorials and continuing bonds. *Mortality* 20(4): 375–389.

boyd D (2006) Friendster lost steam. Is MySpace just a fad?. In: *Apophenia Blog,* 21 March. Available at: http://www.danah.org/papers/FriendsterMySpaceEssay.html (accessed 5 April 2019).

Brown KV (2016) We calculated the year dead people on Facebook could outnumber the living. In: *Fusion.net*, 3 April. Available at: http://fusion.net/story/276237/the-number-of-dead-people-on-facebook-will-soon-outnumber-the-living/ (accessed 1 August 2016).

Brubaker JR and Callison-Burch V (2016) Legacy contact: Designing and implementing post-mortem stewardship at Facebook. In: *Proceedings of the 2016 CHI conference on human factors in computing systems*, San Jose, California, USA, 7–12 May, pp. 2908–2919. Available at: https://doi.org/10.1145/2858036.2858254.

Brubaker JR, Hayes GR and Dourish P (2013) Beyond the grave: Facebook as a site for the expansion of death and mourning. *The Information Society: An International Journal* 29: 152–163. http://doi.org/10.1080/01972243.2013.777300.

Brügger N and Shroeder R (eds) (2017) *The Web as History: Using Web Archives to Understand the Past and the Present*. London: UCL Press.

Cadwalladr C and Graham-Harrison E (2018) Revealed: 50 million Facebook profiles harvested for Cambridge analytica in major data breach. Available at: https://www.theguardian.com/news/2018/mar/17/cambridge-analytica-facebook-influence-us-election (accessed 5 April 2019).

Cameron F and Kenderdine S (2007) *Theorizing Digital Cultural Heritage: A Critical Discourse*. Cambridge, MA: MIT Press.

Cann CK (2014) Tweeting death, posting photos, and pinning memorials: Remembering the dead in bits and pieces. In: Moreman CM, Lewis D (eds) *Digital Death: Mortality and Beyond in the Online Age*. Santa Barbara, CA: Praeger, pp. 69–82.

Choudhary S (2018) A virtual life after death: An exploratory study with special reference to India. *International Journal of Scientific Research in Computer Science, Engineering and Information Technology* 3(3): 1876–1881.

Craig B, Michael A, Martin G, et al. (2013) Consumer Issues for Planning and Managing Digital Legacies. In: *IEEE Technology and Society Magazine*, vol. 33, no. 3, pp. 26–31. doi: 10.1109/MTS.2014.2353751.

Di Cosmo R and Zacchiroli S (2017) Software heritage: Why and how to preserve software source code. In: *iPRES 2017 – 14th international conference on digital preservation*, Kyoto, Japan, September. pp. 1–10.

Facebook (n.d.) Report a deceased person. Available at: https://www.facebook.com/help/408583372511972/ (accessed 24 August 2018).

Facebook (2018) Facebook reports first quarter 2018 results. Available at: https://investor.fb.com/investor-news/press-release-details/2018/Facebook-Reports-First-Quarter-2018-Results/default.aspx (accessed 20 January 2019).

Floridi L (2013) *The Ethics of Information*. Oxford: Oxford University Press.

Fuchs C and Sevignani S (2013) What is digital labour? What is digital work? What's their difference? And why do these questions matter for understanding social media? *TripleC* 11: 237–293.

Gotved S (2014) Research review: Death online – Alive and kicking. *Thanatos* 3(1): 112–126.

Grimm C and Chiasson S (2014) Survey on the fate of digital footprints after death. USEC 14–23 February 2014, San Diego, CA, USA. 2014 Internet Society, ISBN 1-891562-37-1. Available at: http://dx.doi.org/10.14722/usec.2014.23049.

Harbinja E (2014) Virtual worlds – A legal post-mortem account. *SCRIPTed* 11(3). Available at: http://doi.org/10.2966/scrip.110314.273.

Hastie TJ and Tibshirani RJ (1990) *Generalized Additive Models*. Boca Raton, FL: Chapman and Hall/CRC.

Karppi T (2013) Noopolitics of memorializing dead Facebook users. *Culture Machine* 14: 1–20. Available at: www.culturemachine.net.

Kasket E (2012) Being-towards-death in the digital age. *Existential Analysis: Journal of the Society for Existential Analysis* 23(2): 249–261.

Kuny T (1998) A digital dark ages? Challenges in the preservation of electronic information. *International Preservation News* 17(May): 8–13. http://doi.org/Article.

Lavoie B and Dempsey L (2004) Thirteen ways of looking at... digital preservation. *D-Lib Magazine* 10(7–8). http://doi.org/10.1045/july2004-lavoie (accessed 5 April 2019).

Lavoie BF (2003) The incentives to preserve digital materials: Roles, scenarios, and economic decision-making. White paper published electronically by OCLC Research. Available at: https://www.oclc.org/content/dam/research/activities/digipres/incentives-dp.pdf.

Leaver T (2013) The social media contradiction: Data mining and digital death. *M/C Journal* 16(2). http://journal.mediaculture.org.au/index.php/mcjournal/article/viewArticle/625.

Lingel J (2013) The digital remains: Social media and practices of online grief. *The Information Society* 29(3): 190–195. http://doi.org/10.1080/01972243.2013.777311.

Lor PJ and Britz JJ (2012) An ethical perspective on political-economic issues in the long-term preservation of digital heritage Peter. *Journal of the American Society for Information Science and Technology* 63(11): 2153–2164. http://doi.org/10.1002/asi.

Mayer-Schönberger V (2009) *Delete: The Virtue of Forgetting in the Digital Age*. Princeton, NJ: Princeton University Press.

Orwell G (1949) 1984. Available at: https://www.planetebook.com/free ebooks/1984.pdf.

Öhman C (2018) The grand challenges of death in the 21st century. *Swissfuture, Magazin für Zukunftsmonitoring*, 1 May, 16–18.

Öhman C and Floridi L (2017) The political economy of death in the age of information: A critical approach to the digital afterlife industry. *Minds and Machines* 27(4): 639–662. Available at: http://doi.org/10.1007/s11023-017-9445-2 .

Öhman C and Floridi L (2018) An ethical framework for the digital afterlife industry. *Nature Human Behaviour* 2(5): 318–320. Available at:http://doi.org/10.1038/s41562-018-0335-2.

Palm J (2006) *The Digital Black Hole*. Stockholm: Riksarkivet/National Archives.

Pew Research Centre (2018) Social media use continues to rise in developing countries, but plateaus across developed ones. Available at: http://assets.pewresearch.org/wp-content/uploads/sites/2/2018/06/15135408/Pew-Research-Center_Global-Tech-Social-Media-Use_2018.06.19.pdf (accessed 5 April 2019).

Pitsillides S, Jeffries J and Conreen M (2012) Museum of the self and digital death: An emerging curatorial dilemma for digital heritage. In: Giaccardi E (ed.) *Heritage and Social Media: Understanding Heritage in a Participatory Culture*. London/New York, NY: Routledge, pp. 56–68.

R Core Team (2018) *R: A Language and Environment for Statistical Computing*. Vienna: R Foundation for Statistical Computing.

Raymond M (2010) The Library and Twitter: An FAQ. Library of Congress Blog, April 28. Available at: https://blogs.loc.gov/loc/2010/04/the-library-and-twitter-an-faq/ (accessed April 5 2019).

Roland L and Bawden D (2012) The future of history: Investigating the preservation of information in the digital age. *Library & Information History* 28(3): 220–236. http://doi.org/10.1179/1758348912Z.00000000017.

Rothenberg J (1995) Ensuring the longevity of digital documents. *Scientific American* 95(1): 24–29. http://www.clir.org/pubs/archives/ensuring.pdf.

Rubin DB (1981) The Bayesian bootstrap. *Annals of Statistics* 9(1): 130–134. http://doi.org/10.1214/aos/1176345338.

Sherlock A (2013) Larger than life: Digital resurrection and the re-enchantment of society. *The Information Society*

29(3): 164–176. http://doi.org/10.1080/01972243.2013.777302.

Smit E, van der Hoeven J and Giaretta D (2011) Avoiding a digital dark age for data: Why publishers should care about digital preservation. *Learned Publishing* 24(1): 35–49. http://doi.org/10.1087/20110107.

Steinhart E (2007) Survival as a digital ghost. *Minds and Machines* 17(3): 261–271. http://doi.org/10.1007/s11023-007-9068-0.

Stokes P (2012) Ghosts in the machine: DO the dead live on in facebook? *Philosophy and Technology* 25(3): 363–379. http://doi.org/10.1007/s13347-011-0050-7.

Stokes P (2015) Deletion as second death: The moral status of digital remains. *Ethics and Information Technology* 17(4): 1–12. http://doi.org/10.1007/s10676-015-9379-4.

Swan LS and Howard J (2012) Digital immortality: Self or 0010110? *International Journal of Machine Consciousness* 04(01): 245–256. http://doi.org/10.1142/S1793843012400148.

Task Force on Archiving of Digital Information (1996) Report of the task force on archiving of digital information. Available at: https://clir.wordpress.clir.org/wp-content/uploads/sites/6/pub63watersgarrett.pdf (accessed 5 April 2019).

Todd R (2018) Advertiser class action claims Facebook over estimates audience size. *The Recorder*. Available at: https://www.law.com/therecorder/2018/08/16/advertiser-class-action-claims-facebook-over-estimates-audience-size/?fbclid=IwAR0-BE3n8oR1kuzzLFgje7AOKq6D23I6HjaZq4ijO9ziUaB76F8vY-JnniQ (accessed 15 December 2018).

Twitter.com (n.d.) Inactive account policy. Available at: https://help.twitter.com/en/rules-and-policies/inactive-twitter-accounts (accessed 5 April 2019).

United Nations, Department of Economic and Social Affairs (2017) World population prospects: The 2017 revision, custom data acquired via website. World Population Prospects: The 2017 Revision, custom data acquired via website.

Waters D (2002) Good archives make good scholars: Reflections on recent steps toward the archiving of digital information. In: *The state of digital preservation: An international perspective*, Washington, D.C.: Council on Library and Information Resources, pp. 78–95. Available at: http://www.clir.org/pubs/abstract/pub107-abst.html.

Whitt RS (2017) Through a glass, darkly: Technical, policy, and financial actions to avert the coming digital dark ages. *Santa Clara High Techology Law Journal* 33: 117. https://digitalcommons.law.scu.edu/chtlj/vol33/iss2/1.

Wood SN (2006) Low-rank scale-invariant tensor product smooths for generalized additive mixed models. *Biometrics* 62(4): 1025–1036. http://doi.org/10.1111/j.1541-0420.2006.00574.x.

Wood SN (2017) *Generalized Additive Models*, 2nd ed. Boca Raton, FL: Chapman and Hall/CRC.